\documentclass[12pt]{article}
\usepackage{graphicx}
\usepackage{amsmath,amssymb}
\begin{document}
\begin{center}
{\large The sum rule for the structure functions of the
deuteron from the current algebra on the null-plane}\vspace*{2mm}\\
{\large Susumu Koretune} \vspace*{2mm}\\
Department of Physics, Shimane University,
Matsue,Shimane,690-8504,Japan\vspace*{2mm}\\

\end{center}
The fixed-mass sum rules for the deuteron target
have been derived by using the connected matrix element of the
current anti-commutation relation on the null-plane.
From these sum rules we obtain the relation
between the pseudo-scalar meson deuteron total 
cross sections and the structure functions of the deuteron. 
We show that the nuclear effect on the mean hypercharge
of the sea quark of the proton can be studied by this relation.
Further we obtain the relation among the Born term, the resonances 
and the non-resonant background in the small $Q^2$ region,
and as a new aspect of the spin 1 target, 
explicit relation of the tensor structure function
$b_2$ at small or moderate $Q^2$ to that at large $Q^2$.

\section{Introduction}
The current algebra based on the canonical quantization at equal-time
gives us very general constraints.
These constraints are essential ingredient in QCD. \cite{AD}
The sum rules in the current-hadron reaction 
in this formalism are called as the fixed-mass sum rules because the
mass of the current takes the fixed space-like or null value. Adler sum rule and
Adler-Weisberger sum rule are typical examples of these sum rules. In the former
case, the momentum transfer of the weak boson which couple to
the hadronic weak current is fixed at the space-like value, 
and in the latter case the square of the momentum of
the pseudo-scalar meson which
is related to the divergence of the axial-vector current 
is fixed at the off-shell value 0. 
This algebra had been extended to the one based on the canonical 
quantization at equal null-plane time. 
The superior points of the null-plane formalism over the equal-time
one are the followings.(1) We need not take the infinite
momentum frame. (2)Some sum rules 
in the equal-time formalism get corrections
from the bilocal currents, which without them were
considered to be peculiar. (3)A technical
aspect in dealing with some graphs contributing to the
intermediate state.
These were explained in Ref.\cite{DJT}.
Apart from these facts, null-plane formalism involved a
further extension to the current anti-commutation relation.\cite{K80, GER} 
We briefly explain the fact in the following  and technical aspects are
summarized in the Appendix A.\\
In the late 60's, through the experimental finding of the parton at
SLAC, the light-cone current algebra was proposed. This algebra was abstracted
from the leading light-cone singularity of the current commutation
relation in the free quark model \cite{FG}. 
Further, since the leading singularity was
mass-independent, it was suggested that the reasoning to reach 
this algebra could be extended to the current product
if we sacrificed the causal nature of the current commutation relation.
Though this type of the relation had been suggested but 
had not been considered further in Ref.\cite{FG}. 
On the other hand, since the assumption to 
extract the light-cone singularity was too restricted, 
another method which was based on the canonical quantization
on the null-plane was considered. \cite{DJT}  This algebra
was a direct generalization of the equal-time formalism.
The canonical quantization on the null-plane
originated from Dirac\cite{Dir} and was unrelated to the 
light-cone current algebra. However, similar bilocal quantities appeared
in the both methods.
The bilocal quantities in the light-cone current algebra were regular operators
where all singularities in the light-cone limit were taken out
and hence were different from those in the null-plane formalism
where such manipulation had not been imposed. 
However because of the similarity they were often identified 
on the null-plane as a heuristic method to obtain some
physical insight. Among them the works in Ref.\cite{GER} would be a
first attempt to obtain some relations at the wrong signature point
in this sense. 
Now through the finding of the scaling violations which led to the
QCD, it was recognized that the method of taking the leading light-cone
singularity should be refined. In fact, the
short distance expansion was taken first, and with use of 
the dispersion relation, this expansion was analytically continued
to the region near the light-cone.
This light-cone expansion utilized the causality of the current commutation
relation, hence, the moment sum rules obtained 
in this expansion were at alternate
integers. Further, each moment corresponded to the matrix element of the 
local operator obtained by the expansion of
the bilocal operator and it was found that because of the anomalous dimension we
could not take out the light-cone singularity uniformly from each
moment. Thus the expansion by the singular coefficient function
multiplied by the regular bilocal current in the light-cone current
algebra had broken down. 
The relation at the missing integers was later shown to 
be obtained by the cut vertex formalism\cite{Mul} which suggested that
these quantities were related to the non-local quantities. A physical
application of the light-cone expansion were restricted to the deep-inelastic region.\\
Now through the study of the fixed-mass sum rules in the semi-inclusive
lepton-hadron scatterings where the one soft pion was observed,
we encountered the current anti-commutation relation on the null-plane.
Since at that time we knew the simple method to take out the leading
light-cone singularity was wrong and the bilocal operator 
in this method should not be taken literally, we needed some methods
to abstract the current anti-commutation relation on the null-plane.
It was in this point where Deser, Gilbert and Sudarshan (hereafter called
as DGS) representation\cite{DGS} played an important role.\cite{K80} 
Through this method it became possible to consider 
the fixed-mass sum rules at the wrong signature point with use of the
connected matrix element of the current anti-commutation relation 
between the stable hadron. The application of this method has been so far
restricted to the hadrons. However, as far as the $s$ channel and the
$u$ channel are disconnected and that the target particle is stable, 
this method can be used.  Now the
sum rules from the current anti-commutation relation gave us information
of the sea quarks in the hadrons. A typical example of this fact can
be seen in the modified Gottfried sum rule.\cite{K93,K95,K97,K98} 
Compared with the Adler sum rule which is obtained 
by the current commutation relation,
we have the extra factor $(-1)$ from the contribution of the anti-quark
distributions.\cite{K98}  Hence the contribution from the sea quark distribution
remains in the sum rule. Thus the study of the sum
rules can give us information of the hadronic vacuum.  In other words,
we can say that the sum rule controls how the quark-antiquark pair is produced
or annihilated in the hadrons. From this point of view, it is
interesting to extend the method to the nuclear target since an nuclear effect
is different from that of the hadron.  In this paper, as a first step
of the application to the nuclear targets, we apply the method to the deuteron. 
In section 2, the kinematics of the spin 1 deuteron target is given, 
and in section 3,
the sum rules from the good-good component are derived. In section 4,
the sum rules are transformed to various forms and physical meanings
are explained. Summary is given in Section 5.

\section{Kinematics}
The imaginary part of the forward reaction ''current(q) +
deuteron(p) $\to$ current(q) + deuteron(p)'' is
proportional to the total cross section of the inclusive
reaction ''current(q) + deuteron(p) $\to$ anythings(X)'', where $q$ is the
momentum of the current and $p$ is that of the deuteron with its mass $m_d$. 
This part is called as the hadronic tensor and is expressed
by assuming the completeness of the sum over $X$ as
\begin{equation}
W^{\mu \nu}_{ab}(p,q,E,E^{*})
= \frac{1}{4\pi}\int d^{4}x\mathrm{e}^{iq\cdot x}\langle p,E|
[ J_a^{\mu}(x),J_b^{\nu}(0) ] |p,E\rangle_{c},
\end{equation}
where $E$ is the polarization vector of the deuteron and the suffix $c$
on the right-hand side of the equation means to take the connected
part.  Since the current
is the induced hadronic current in the inclusive reaction ''lepton + deuteron $\to$
lepton + anythings'', the momentum $q$ is the difference of the momentum of the initial
lepton and that of the final lepton and hence
it takes the space-like value.  We first discuss the conserved vector 
current  $J_a^{\mu}(x)$ where the suffice $a$ denotes the flavor index. The generalization to the non-conserved
and the parity violating case is given later in this section.
Since the hadronic current is color singlet we ignore a color suffix in the
quark field. Now, by requiring parity and time reversal invariance, we obtain\cite{HJM}
\begin{eqnarray}
W^{\mu \nu}_{ab}(p,q,E,E^{*})
=-F_1^{ab}G^{\mu\nu} +F_2^{ab}\frac{P^{\mu}P^{\nu}}{\nu}
-b_1^{ab}r^{\mu\nu}+
\frac{1}{6}b_2^{ab}(s^{\mu\nu}+t^{\mu\nu}+u^{\mu\nu})\nonumber \\
+\frac{1}{2}b_3^{ab}(s^{\mu\nu}-u^{\mu\nu})
+\frac{1}{2}b_4^{ab}(s^{\mu\nu}-t^{\mu\nu})
+\frac{ig_1^{ab}}{\nu}\epsilon^{\mu\nu\lambda\sigma}q_{\lambda}s_{\sigma}
+\frac{ig_2^{ab}}{\nu^2}\epsilon^{\mu\nu\lambda\sigma}q_{\lambda}
(\nu s_{\sigma}-s\cdot qp_{\sigma}),
\end{eqnarray}
where $\nu =p\cdot q$ , $\kappa = 1 - m_d^2q^2/\nu^2$,
$P^{\mu}=p^{\mu}-(\nu /q^2)q^{\mu},G^{\mu\nu}=g^{\mu\nu}- (1/q^2)q^{\mu}q^{\nu},$
$s^{\mu}=-(i/m_d^2)\epsilon^{\mu\alpha\beta\gamma}E^{*}_{\alpha}E_{\beta}p_{\gamma}$,
$E\cdot E^{*}=-m_d^2$ , $p\cdot E = p\cdot E^{*} = 0$ , and
\begin{eqnarray}
r^{\mu\nu}&=&\frac{1}{\nu^2}(q\cdot E^{*}q\cdot E - \frac{1}{3}\nu^2\kappa)G^{\mu\nu},\hspace*{1cm}
s^{\mu\nu}=\frac{2}{\nu^2}(q\cdot E^{*}q\cdot E - \frac{1}{3}\nu^2\kappa)\frac{P^{\mu}P^{\nu}}{\nu},\nonumber \\
t^{\mu\nu}&=&\frac{1}{2\nu^2}(q\cdot E^{*}P^{\mu}\widetilde{E}^{\nu}+
q\cdot E^{*}P^{\nu}\widetilde{E}^{\mu}+q\cdot EP^{\mu}\widetilde{E}^{*\nu}
+q\cdot EP^{\nu}\widetilde{E}^{*\mu}-\frac{4\nu}{3}P^{\mu}P^{\nu}),\nonumber \\
u^{\mu\nu}&=&\frac{1}{\nu}(\widetilde{E}^{*\mu}\widetilde{E}^{\nu}+
\widetilde{E}^{*\nu}\widetilde{E}^{\mu}+\frac{2m_d^2}{3}G^{\mu\nu}-\frac{2}{3}P^{\mu}P^{\nu}),
\end{eqnarray}
with $\widetilde{E}^{\mu}=E^{\mu}-(q\cdot
E/q^2)q^{\mu}$ and $\widetilde{E}^{*\mu}=E^{*\mu}-(q\cdot
E^{*}/q^2)q^{\mu}$.\\
Similar hadronic tensor $\widetilde{W}^{\mu\nu}_{ab}(p,q,E,E^{*})$
can be defined by the current anti-commutation relation as
\begin{eqnarray}
\widetilde{W}^{\mu \nu}_{ab}(p,q,E,E^{*})
= \frac{1}{4\pi}\int d^{4}x\mathrm{e}^{iq\cdot x}\langle p,E|
\{ J_a^{\mu}(x),J_b^{\nu}(0)\}|p,E\rangle_{c} \nonumber \\
= -\widetilde{F}_1^{ab}G^{\mu\nu} +\widetilde{F}_2^{ab}\frac{P^{\mu}P^{\nu}}{\nu}
-\widetilde{b}_1^{ab}r^{\mu\nu}+
\frac{1}{6}\widetilde{b}_2^{ab}(s^{\mu\nu}+t^{\mu\nu}+u^{\mu\nu})+\frac{1}{2}\widetilde{b}_3^{ab}(s^{\mu\nu}-u^{\mu\nu})\nonumber \\
+\frac{1}{2}\widetilde{b}_4^{ab}(s^{\mu\nu}-t^{\mu\nu})
+\frac{i\widetilde{g}_1^{ab}}{\nu}\epsilon^{\mu\nu\lambda\sigma}q_{\lambda}s_{\sigma}
+\frac{i\widetilde{g}_2^{ab}}{\nu^2}\epsilon^{\mu\nu\lambda\sigma}q_{\lambda}
(\nu s_{\sigma}-s\cdot qp_{\sigma}).
 \end{eqnarray}
The structure functions defined by the current commutation relation and
those of the current anti-commutation relation are the same quantity
in the $s$ channel but opposite in sign in the $u$ channel. The crossing relation under 
$\mu \leftrightarrow \nu$ , $a \leftrightarrow b$ and $q\to -q$, are
$F_1^{ab}(-x,Q^2)=-F_1^{ba}(x,Q^2),F_2^{ab}(-x,Q^2)=F_2^{ba}(x,Q^2),
b_1^{ab}(-x,Q^2)=-b_1^{ba}(x,Q^2)$,\\
$b_2^{ab}(-x,Q^2)=b_2^{ba}(x,Q^2)$,
$b_3^{ab}(-x,Q^2)=b_3^{ba}(x,Q^2),b_4^{ab}(-x,Q^2)=b_4^{ba}(x,Q^2),$
$g_1^{ab}(-x,Q^2)=g_1^{ba}(x,Q^2),g_2^{ab}(-x,Q^2)=g_2^{ba}(x,Q^2)$,
while the structure functions defined by the anti-commutation relation
are opposite in sign, where $x=Q^2/2\nu$ with $q^2=-Q^2$.\\
Now we take the current as
$J^{\mu}_a(x)=:\bar{q}(x)\gamma^{\mu}\frac{\lambda_a}{2}q(x):$ 
in the chiral $SU(N)\times SU(N)$ model. On the null-plane $x^+=0$,
the quark field is decomposed as $q^{(\pm )}(x)=\Lambda^{\pm}q(x)$
where the projection operator is define as 
$\Lambda^{\pm}=\frac{1}{2}(1\pm \gamma^0\gamma^3)$ with 
$x^{\pm}=\frac{1}{\sqrt{2}}(x^0\pm x^3)$ and the suffixes of
internal symmetry are discarded since inclusion of them do not
affect the following discussion.  Through the equation of
motion, $q^{(-)}(x)$ is related to the $q^{(+)}(x)$.
Hence the $q^{(-)}(x)$ depends on the $q^{(+)}(x)$, and the independent
field on the null-plane is $q^{(+)}(x)$, hence the canonical quantization
is given as
$\{q^{(+)\dagger}(x),q^{(+)}(0)\}|_{x^+=0}=\sqrt{2}\Lambda^{+}\delta^2(\vec{x}^{\bot})\delta({x^-})$. 
Since
$J^{+}_a(x)=:\bar{q}(x)\gamma^{+}\frac{\lambda_a}{2}q(x):=\sqrt{2}:q^{(+)\dagger}(x)\frac{\lambda_a}{2}q^{(+)}(x):$,
the current for $\mu =+$ depends only on the $q^{(+)}(x)$, and does not
depend on the equation of motion. In this sense the current commutation relation on the null-plane
for $\mu = \nu = +$ is called as the good-good component. The current
$J_a^{i}(x)$ depends on one $q^{(-)}(x)$ and then called as a bad
component.  Thus the good-good component 
is
\begin{eqnarray}
\lefteqn{<p,E|[J^{+}_{a}(x), J^{+}_{b}(0)]|p,E>_c|_{x^+=0}}&& \nonumber \\
&=& i\delta (x^-)\delta^2
 (\vec{x}^{\bot })\left[d_{abc}<p,E|A_c^{+}(x|0)|p,E>_c + f_{abc}<p,E|S_c^{+}(x|0)]|p,E>_c \right],
\end{eqnarray}
where
\begin{eqnarray}
S_a^{\mu}(x|0)&=&
\frac{1}{2}[:\bar{q}(x)\gamma^{\mu}\frac{\lambda_a}{2}q(0):
+ :\bar{q}(0)\gamma^{\mu}\frac{\lambda_a}{2}q(x):] ,\nonumber \\
A_a^{\mu}(x|0)&=&
\frac{1}{2i}[:\bar{q}(x)\gamma^{\mu}\frac{\lambda_a}{2}q(0):
- :\bar{q}(0)\gamma^{\mu}\frac{\lambda_a}{2}q(x):] ,\nonumber \\
\lefteqn{<p,E|S_a^{\mu}(x|0)|p,E>_c
=p^{\mu}S_a(p\cdot x , x^2) +
x^{\mu}\bar{S}_a(p\cdot x , x^2) }&&\nonumber \\
&+& p^{\mu}\{(E^{*}\cdot x)
(E\cdot x) - \frac{1}{3}((x\cdot p)^2-m_d^2x^2)\}S_a^{P}(p\cdot x , x^2)\nonumber \\
&+&x^{\mu}\{(E^{*}\cdot x)
(E\cdot x) - \frac{1}{3}((x\cdot p)^2-m_d^2x^2)\}\bar{S}_a^{P}(p\cdot x, x^2)\nonumber \\
&+&\{ E^{\mu}(E^{*}\cdot x)+E^{*\mu}(E\cdot x)-\frac{2}{3}((x\cdot p)p^{\mu}
-m_d^2x^{\mu})\}\widetilde{S}_a^{P}(p\cdot x , x^2),\nonumber \\
\lefteqn{<p,E|A_a^{\mu}(x|0)|p,E>_c
=p^{\mu}A_a(p\cdot x , x^2) +
x^{\mu}\bar{A}_a(p\cdot x , x^2)}&& \nonumber \\
&+& p^{\mu}\{(E^{*}\cdot x)
(E\cdot x) - \frac{1}{3}((x\cdot p)^2-m_d^2x^2)\}A_a^{P}(p\cdot x , x^2)\nonumber \\
&+& x^{\mu}\{(E^{*}\cdot x)
(E\cdot x) - \frac{1}{3}((x\cdot p)^2-m_d^2x^2)\}\bar{A}_a^{P}(p\cdot x, x^2)\nonumber \\
&+&\{ E^{\mu}(E^{*}\cdot x)+E^{*\mu}(E\cdot x)-\frac{2}{3}((x\cdot p)p^{\mu}
-m_d^2x^{\mu})\}\widetilde{A}_a^{P}(p\cdot x , x^2).
\end{eqnarray}
The target polarization dependent parts are defined so that
their contributions vanish when the target polarizations are averaged.
 The right-hand side of Eq.(5) is equal to $i\delta (x^-)\delta^2 (\vec{x}^{\bot})
f_{abc}<p,E|J_c^+(0)|p,E>_c$ because of the delta function
constraint, however, we write the expression before this constraint is applied,
since what the DGS representation gives us is that the term
corresponding to this term remains in the
anti-commutation relation and that the other terms are zero on the null-plane.\cite{K80,K93}
Then the corresponding relation for the current anti-commutation relation is
\begin{eqnarray}
\lefteqn{<p,E|\{J_a^+(x),J_b^+(0)\}|p,E>_c|_{x^+=0}}&\nonumber \\
=& \frac{1}{\pi }P(\frac{1}{x^-})\delta^2(\vec{x}^{\bot })
\left[d_{abc}<p,E|A_c^{+}(x|0)|p,E>_c + f_{abc}<p,E|S_c^{+}(x|0)]|p,E>_c \right],
\end{eqnarray}
where $P$ means to take the principal value.
Before going to a detailed derivation of the sum rule we explain
the hadronic tensor for the non-conserved and parity violating currents
including the cases for the weak boson mediated reactions.
In such a general case, we have the 36 independent helicity amplitudes
since we have two types of the helicity 0 state for the non-conserved current.\cite{Ji}
Then by the time reversal invariance the independent amplitudes are reduced to 24 and
by the parity invariance they are further reduced to 14.
Among the 14 amplitudes, 6 amplitudes enter due to the 
non-conservation of the current. The tensors corresponding to these
amplitudes are
\begin{eqnarray}
& p^{\mu}q^{\nu}+p^{\nu}q^{\mu}, \quad  (q\cdot E^{*}q\cdot E -
 \frac{1}{3}\nu^2\kappa)(p^{\mu}q^{\nu}+p^{\nu}q^{\mu}), \quad 
(q\cdot E^{*}q\cdot E - \frac{1}{3}\nu^2\kappa)q^{\mu}q^{\nu}, \quad 
q^{\mu}q^{\nu}, \nonumber \\
&\left\{q\cdot E^{*}q^{\mu}\widetilde{E}^{\nu}+
q\cdot E^{*}q^{\nu}\widetilde{E}^{\mu}+q\cdot Eq^{\mu}\widetilde{E}^{*\nu}
+q\cdot Eq^{\nu}\widetilde{E}^{*\mu} - \frac{2}{3}\nu
(p^{\mu}q^{\nu}+p^{\nu}q^{\mu})
+\frac{4\nu^2}{3q^2}q^{\mu}q^{\nu}\right\}, \nonumber \\
\lefteqn{\qquad  i\epsilon^{\mu\nu\alpha\beta}p_{\alpha}s_{\beta}.}&
\end{eqnarray}
\section{Sum rules from the good-good component}
Now, with use of Eqs.(5)-(7), the sum rule can be obtained by 
integrating $W^{++}_{ab}$ and $\widetilde{W}^{++}_{ab}$ over $q^-$ and
assuming the interchange of setting $q^+=0$ and $\nu$ integration.
For the polarization averaged quantities, we obtain from
the current commutation relation
\begin{equation}
\int_0^1\frac{dx}{x}F_2^{[ab]}(x,Q^2) = \frac{1}{4}f_{abc}\Gamma_c,
\end{equation}
where $F_2^{[ab]}=(F_2^{ab}-F_2^{ba})/2i$ and
$<p,E|J_a^{\mu}(0)|p,E>=p^{\mu}\Gamma_a$, and from
the current anti-commutation relation
\begin{equation}
\int_0^1\frac{dx}{x}F_2^{(ab)}(x,Q^2) = \frac{1}{4\pi}d_{abc}
P\int_{-\infty}^{\infty}\frac{d\alpha}{\alpha}A_c(\alpha ,0),
\end{equation}
where $F_2^{(ab)}=(F_2^{ab}+F_2^{ba})/2$ . These sum rules
take the same form as in the case of the nucleon target.
Let us now derive the sum rules for the polarization dependent part.
We denote the helicity of the polarization vector as $E_{h}$, and
take $p=(p^0,0,0,p^3),
q=(q^0,q^1,q^2,q^3),\sqrt{2}E_{\pm}=m_d(0,\mp,-i,0),E_0=(p^3,0,0,p^0)$.  
Then, by taking the polarization vector $E_{\pm}$, 
we obtain from the commutation relation
\begin{equation}
\int_0^1\frac{dx}{x}\left( (\kappa -7)b_2^{[ab]}(x,Q^2) +3(\kappa -1)b_3^{[ab]}(x,Q^2)
+3(\kappa -1)b_4^{[ab]}(x,Q^2)\right) = 0,
\end{equation}
and from the anti-commutation relation
\begin{eqnarray}
\int_0^1\frac{dx}{x}\left( (\kappa -7)b_2^{(ab)}(x,Q^2) +3(\kappa -1)b_3^{(ab)}(x,Q^2)
+3(\kappa -1)b_4^{(ab)}(x,Q^2)\right)\nonumber \\
= -\frac{3}{2\pi}d_{abc}
\int_{-\infty}^{\infty}d\alpha \{\alpha A_c^{P}(\alpha,0)+2\widetilde{A}_c^{P}(\alpha ,0)\} ,
\end{eqnarray}
where symmetric and antisymmetric combination of the spin dependent
structure functions are defined similarly as the structure
function $F_2^{ab}$. From the case $E_0$, we obtain no new sum rules.
Now as the transverse polarization vector, we take the combination
$E_1=m_d(0,1,0,0)$ and $E_{2}=m_d(0,0,1,0)$, and $p=(p^0,0,0,p^3),
q=(q^0,q^1,0,q^3)$. Then, since $E_2\cdot q=E_2^{*}\cdot q=0$,
we obtain from the commutation relation 
\begin{equation}
\int_0^1\frac{dx}{x}\left((\kappa + 2)b_2^{[ab]}(x,Q^2) +3(\kappa -1)b_3^{[ab](x,Q^2)}
+3(\kappa -1)b_4^{[ab]}(x,Q^2)\right) =0 ,
\end{equation}
and from the anti-commutation relation
\begin{eqnarray}
\int_0^1\frac{dx}{x}\left( (\kappa +2)b_2^{(ab)}(x,Q^2) +3(\kappa -1)b_3^{(ab)}(x,Q^2)
+3(\kappa -1)b_4^{(ab)}(x,Q^2)\right) \nonumber \\
= \frac{3}{4\pi}d_{abc}
\int_{-\infty}^{\infty}d\alpha \{\alpha A_c^{P}(\alpha
,0)+2\widetilde{A}_c^{P}(\alpha ,0)\} .
\end{eqnarray}
Since $\kappa -1=4x^2m_d^2/Q^2$, from Eq.(11) and Eq.(13)
we obtain
\begin{equation}
\int_0^1\frac{dx}{x}b_2^{[ab]}(x,Q^2) = 0 ,
\end{equation}
and
\begin{equation}
\int_0^1dx x\left(b_2^{[ab]}(x,Q^2)+3b_3^{[ab]}(x,Q^2)+3b_4^{[ab]}(x,Q^2)\right)=0.
\end{equation}
Similarly, from Eq.(12) and Eq.(14) we obtain
\begin{equation}
\int_0^1\frac{dx}{x}b_2^{(ab)}(x,Q^2)=
\frac{1}{4\pi}d_{abc}\int_{-\infty}^{\infty}d\alpha \{\alpha
A_c^{P}(\alpha,0)
+2\widetilde{A}_c^{P}(\alpha ,0)\} ,
\end{equation}
and
\begin{equation}
\int_0^1dx x\left(b_2^{(ab)}(x,Q^2)+3b_3^{(ab)}(x,Q^2)
+3b_4^{(ab)}(x,Q^2)\right) =0.
\end{equation}
The sum rules (10) ,(17) and (18) are the ones for the symmetric
combination, hence they can be applied to the electromagnetic
current. \\
Now, since \\
$<p,E|[J_a^{5+}(x),J_b^{5+}(0)]|p,E>_c|_{x^+=0} =
<p,E|[J_a^+(x),J_b^+(0)]|p,E>_c|_{x^+=0} $, \\
we obtain the relation\\
 $<p,E|\{J_a^{5+}(x),J_b^{5+}(0)\}|p,E>_c|_{x^+=0} =
<p,E|\{J_a^+(x),J_b^+(0)\}|p,E>_c|_{x^+=0}$ with use of the
DGS representation\cite{K80,K93}. The hadronic tensor for
the axial-vector current is non-conserved one. Then
the tensors given in Eq.(8) are necessary. 
When we derive the sum rule we set $q^+=0$. Since all tensors
in Eq.(8) are proportional to $q^+$, we see that they do not affect
the derivation of the sum rule in the above discussion. Thus 
the sum rule (10) also holds in this case. Then by using the
PCAC relation, we can transform the sum rule (10) to the
ones for the pseudo-scalar deuteron total cross section
as in the nucleon case\cite{K80,K93}. 

\section{Application}
Now, the sum rule (10) is the equality of the possible
divergent quantity which definitely breaks the condition
necessary to derive the sum rule. Including such case, 
importance of the regularization of possible divergent sum rules was
explained in Ref.\cite{BFJ}. Here we follow the method in Ref.\cite{DeAlwis}.
We first derive the sum rule in the non-forward direction. 
Then we see that the right-hand side of the sum rule
given by the integral of the non-forward matrix element of the 
bilocal current is $Q^2$ independent. We assume the high-energy behavior
is controlled by the moving Regge pole or cut and the divergence comes
from the flavor singlet part corresponding to the Pomeron. Then we take sufficiently
large $|t|$ such that the sum rule is convergent. 
Next we change $|t|$ to smaller value, and subtract the pole singularity
from both-hand sides of the sum rule. From this we obtain the condition
that the residue of the pole is $Q^2$ independent. 
After that, we can take still smaller $|t|$ in
the subtracted sum rule, and we finally obtain the relation at $t=0$.
A net result of this manipulation can be mimicked in the forward sum
rule by changing the intercept of the slope parameter
appropriately. \cite{K93,K95,K97, K98}
Now we take the chiral $SU(3)\times SU(3)$ flavor symmetry and
obtain
\begin{equation}
B_{\pi}+\frac{2f_{\pi}^2}{\pi}\int_{\nu_0^{\pi}}^{\infty}\frac{d\nu}{\nu}
\{\sigma^{\pi^{+}d}(\nu )+\sigma^{\pi^{-}d}(\nu )\}
=\frac{1}{2\pi}P\int_{-\infty}^{\infty}\frac{d\alpha}{\alpha}
\{\frac{2\sqrt{6}}{3}A_0(\alpha ,0)+ \frac{2\sqrt{3}}{3}A_8(\alpha ,0)\},
\end{equation}
\begin{equation}
B_{K}+\frac{2f_{K}^2}{\pi}\int_{\nu_0^{K}}^{\infty}\frac{d\nu}{\nu}
\{\sigma^{K^{+}d}(\nu )+\sigma^{K^{-}d}(\nu )\}
=\frac{1}{2\pi}P\int_{-\infty}^{\infty}\frac{d\alpha}{\alpha}
\{\frac{2\sqrt{6}}{3}A_0(\alpha ,0)+A_3(\alpha ,0) - \frac{\sqrt{3}}{3}A_8(\alpha ,0)\},
\end{equation}
where $\sigma$ means the off-shell $q^2=0$ total cross section of the reaction
specified by its upper suffix and can be assumed to be smoothly
continued to the on-shell one, and  $\nu_0^{\pi}=m_{\pi}m_d$ and
$\nu_0^{K}=m_{K}m_d$. The $f_{\pi}$ and the $f_{K}$ are
the pion and the kaon decay constant respectively. $B_{\pi}$ and 
$B_{K}$ is the contribution from the Born terms and the unphysical
region below the threshold of the continuum contribution.
In the neutrino reactions, we obtain
\begin{equation}
\int_0^1\frac{dx}{2x}\{F_2^{\bar{\nu}d}(x,Q^2)+F_2^{\nu d}(x,Q^2)\}
=\frac{1}{2\pi}P\int_{-\infty}^{\infty}\frac{d\alpha}{\alpha}
\{\frac{2\sqrt{6}}{3}A_0(\alpha ,0)+ \frac{2\sqrt{3}}{3}A_8(\alpha ,0)\},
\end{equation}
and in the electroproduction we obtain   
\begin{equation}
\int_0^1\frac{dx}{x}F_2^{ed}(x,Q^2) = \frac{1}{18\pi}
P\int_{-\infty}^{\infty}\frac{d\alpha}{\alpha}
\{2\sqrt{6}A_0(\alpha ,0)+3A_3(\alpha ,0) +\sqrt{3}A_8(\alpha ,0)\}.
\end{equation}
In the left-hand side of the sum rules (21) and (22),
the contribution from the Born term is included but can be neglected
in the deep-inelastic region. 
We regularize the sum rules (19) - (22) by the method
explained just before Eq.(19)\cite{K93,K95,K97, K98}.
In addition here we assume
$\displaystyle{P\int\frac{d\alpha}{\alpha}A_3(\alpha ,0)=0}$
since the deuteron is iso-singlet. Note that this quantity corresponds to the
difference of the mean $I_3$ of the quark and the anti-quark in the
proton and that in the neutron in the deuteron, and hence it is zero
under the isospin symmetry. In this way, we obtain the relation 
\begin{equation}
\int_{0}^{1}\frac{dx}{x}\left\{\left(\frac{F_2^{\bar{\nu}d}(x,Q^2)+F_2^{\nu d}(x,Q^2)}{2}
\right)-3F_2^{ed}(x,Q^2)\right\} = \frac{I_{\pi}^d - I_{K}^d}{3},
\end{equation}
where, by assuming the smooth extrapolation to the on-shell quantity, 
$I_{\pi}$ and $I_K$ are defined as
\begin{equation}
I_{\pi}^d =B_{\pi}+\frac{2f_{\pi}^2}{\pi}\int_{\nu_0^{\pi}}^{\infty}\frac{d\nu}{\nu^2}
\left[(\nu^2-m_{\pi}^2m_d^2)^{1/2}\{\sigma^{\pi^{+}d}(\nu
)+\sigma^{\pi^{-}d}(\nu )\} -\nu s^b_{\pi}\beta_{\pi d}\right]
+\frac{2f_{\pi}^2\beta_{\pi
N}}{\pi}\ln{\left[\frac{1}{2\nu_0^{\pi}}\right]},
\end{equation}
\begin{equation}
 I_{K}^d =B_{K}+\frac{2f_{K}^2}{\pi}\int_{\nu_0^{K}}^{\infty}\frac{d\nu}{\nu^2}
\left[(\nu^2-m_{K}^2m_d^2)^{1/2}\{\sigma^{K^{+}d}(\nu )+\sigma^{K^{-}d}(\nu
)\}-\nu s^b_{K}\beta_{K d}\right]
+\frac{2f_{K}^2\beta_{KN}}{\pi}\ln{\left[\frac{1}{2\nu_0^{K}}\right]},
\end{equation}
with the leading high energy behavior being given by the soft Pomeron \cite{LD}as
\begin{equation}
\left\{\sigma^{\pi^{+}d}(\nu )+\sigma^{\pi^{-}d}(\nu )\right\}\sim \beta_{\pi
 d}s_{\pi}^{\alpha_P(0)-1},\quad
\left\{\sigma^{K^{+}d}(\nu )+\sigma^{K^{-}d}(\nu )\right\}\sim \beta_{K
 d}s_{K}^{\alpha_P(0)-1},
\end{equation}
where $\alpha_P(0)=1+b$ with $b=0.0808$, $s_{\pi}=m_{\pi}^2+m_d^2+2\nu$ and
$s_{K}=m_{K}^2+m_d^2+2\nu$, and as a result of the assumption 
that the divergence in the forward
direction comes from the singlet, we obtain 
$f_{\pi}^2\beta_{\pi d}=f_{K}^2\beta_{K d}$, and 
the relation between the residue of the Pomeron in the pion deuteron
cross section and that of the structure function in the lepton-hadron
scatterings.\cite{K93,K95,K97, K98}
In terms of the sea quark distribution function $\lambda_i(x,Q^2)$of the
proton in the 
deuteron where $i=u,d,s$ specifys the quark, the sum rule (23) can be transformed as
\begin{equation}
\frac{1}{3}\int_0^1dx\left\{\lambda_u(x,Q^2)+\lambda_d(x,Q^2)
-2\lambda_s(x,Q^2)\right\}
=\frac{1}{2}\left( -1 + \frac{I_{\pi}^d - I_{K}^d}{3} \right).
\end{equation}
In Eq.(27), we have assumed the isospin symmetry of the quark
distribution function and expressed
the quark distribution function of the neutron in the deuteron
by the one of the proton in the deuteron. Further, though
we take $\lambda_{\bar{i}}=\lambda_i$ for simplicity, what is really
required in our formalism is the equality of the integrated quantity. Hence
they can take different value locally.
The left-hand side of Eq.(27) is the mean hypercharge of the sea quark
of the proton in the deuteron. If we neglect the nuclear effect, 
the deuteron cross section is the sum of that of the proton and the neutron.
In this case we obtain the relations 
$I_{\pi}^d\approx 2I_{\pi}$ and $I_{K}^d\approx I_{K}^{p}+I_{K}^n$.
Using these relations on the right-hand side of Eq.(27),
we find that the left-hand side of it exactly agrees with the mean hypercharge of the 
sea quark of the proton given in Refs.\cite{K97, K98}. Since these
relations break down due to the nuclear effects,
we see the sum rule (23) or (27) gives us information of the hadronic
vacuum under the nuclear environment. Now, in a phenomenological analysis,
we may need modification of the sum rules (23) and (27). These sum rules
are derived by the assumption where the Pomeron is flavor singlet.
The condition $f_{\pi}^2\beta_{\pi d}=f_{K}^2\beta_{K d}$
obtained by this assumption is violated phenomenologically.
One way to account this effect is explained in Ref.\cite{K95}. 
However, it should be noted that the sum rules
(23) and (27) correspond to the quantity related to the 
hypercharge and hence have a clear
physical meaning. They show that the
large symmetry restoration of the strange sea quark is necessary in the small
$x$ region. Since the strange sea quark distribution is suppressed
above $x=0.01$ greatly, this symmetry restoration itself is an
interesting phenomena. Hence, these relation should be studied first by neglecting
the symmetry breaking effect and taking the
symmetry limit of sea quark distributions.\cite{K03} We explain the
possible symmetry breaking effects together with the symmetry relation
in the Appendix B. \\
Another application of Eq.(22) is to consider the relation for arbitrary two
different $Q_1^2$ and $Q_2^2$ by separating out the Born term from $F_2^{ed}$.
\begin{equation}
\int_{x_c(Q_1^2)}^1\frac{dx}{x}F_2^{ed}(x,Q_1^2)
-\int_{x_c(Q_2^2)}^1\frac{dx}{x}F_2^{ed}(x,Q_2^2)
=B(Q_1^2,Q_2^2) + K^{ed}(Q_1^2,Q_2^2),
\end{equation} 
where the contribution from the Born term is given as
\begin{equation}
B(Q_1^2,Q_2^2)=\left[G_C^2(Q_2^2)+\frac{8}{9}\eta_2^2G_Q^2(Q_2^2)
+\frac{2}{3}\eta_2G_M^2(Q_2^2)\right]
-\left[G_C^2(Q_1^2)+\frac{8}{9}\eta_1^2G_Q^2(Q_1^2)
+\frac{2}{3}\eta_1G_M^2(Q_1^2)\right] ,
\end{equation}
with $\eta_i=Q_i^2/4m_d^2$.  $G_C , G_M $ and $G_Q$ are charge, magnetic
and quadrupole moment of the deuteron defined as
\begin{eqnarray}
\lefteqn{<n,E^{\prime}|J^{\mu}_{em}(0)|p,E>}&\nonumber \\
=&-\frac{1}{m_d^2}\Big(\big\{G_1(Q^2)(E^{\prime *}\cdot E) -G_3(Q^2)\frac{(E^{\prime
 *}\cdot q)(E\cdot q)}{2m_d^2}\big\}(p+n)^{\mu} \nonumber \\
&+ G_M(Q^2)\big\{E^{\mu}(E^{\prime
 *}\cdot q)-E^{\prime * \mu}(E\cdot q)\big\}\Big)
\end{eqnarray}
with $q=n-p$
for the electromagnetic current
$J^{\mu}_{em}(0)$, and $G_1$ and $G_3$ are related to $G_C , G_M $ and
$G_Q$ as $G_C=G_1+\frac{2}{3}\eta G_Q, G_Q=G_1-G_M+(1+\eta )G_3$
with $\eta = Q^2/4m_d^2$. The derivation of the Born term is
straightforward but tedious hence we give its sketch in the Appendix C.
$K^{ed}(Q_1^2,Q_2^2)$ is defined as
\begin{equation}
K^{ed}(Q_1^2,Q_2^2)=-\int^{x_c(Q_1^2)}_0\frac{dx}{x}F_2^{ed}(x,Q_1^2)
+\int^{x_c(Q_2^2)}_0\frac{dx}{x}F_2^{ed}(x,Q_2^2),
\end{equation}
where $x_c(Q^2)=Q^2/2\nu_c(Q^2)$ with $\nu_c(Q^2)=(W_c^2-m_d^2+Q^2)/2$ . 
Here we define $W^2=(p+q)^2$ and
$W_c$ is the cutoff invariant mass $W$.
In Eq.(31), the integral over $x$ should be taken after subtracting the small
$x$ behavior of $F_2^{ed}(x,Q_1^2)$ and $F_2^{ed}(x,Q_2^2)$
by obtaining the condition that the residue of the pole is $Q^2$
independent. It should be noted that , in this regularization, we need not
consider the symmetry breaking effect of the Pomeron.
In these sum rules, we take $Q_1^2$ fixed and 
can investigate the $Q_2^2$ dependence
of the sum rule.
Further, if we take $Q_1^2$ and $Q_2^2$ small
value such that $K^{ed}(Q_1^2,Q_2^2)$ is negligibly small,
we have the relation which express an intimate relation among
the Born term, the resonances and the non-resonant background.

The regularization of the sum rule (17) can be done similarly, and
we can transform it to
\begin{equation}
\int_{x_c(Q_1^2)}^1\frac{dx}{x}b_2^{ed}(x,Q_1^2)
-\int_{x_c(Q_2^2)}^1\frac{dx}{x}b_2^{ed}(x,Q_2^2)
=B_{b2}(Q_1^2,Q_2^2) + K_{b2}^{ed}((Q_1^2,Q_2^2),
\end{equation}
where
\begin{eqnarray}
B_{b2}(Q_1^2,Q_2^2)&=&4\eta_2\left[\frac{\eta_2}{1+\eta_2}\{G_C(Q_2^2)
+\frac{\eta_2}{3}G_Q(Q_2^2)-G_M(Q_2^2)\}G_Q(Q_2^2)+\frac{1}{4}G_M^2(Q_2^2)\right]\nonumber \\
&\hspace*{-20mm}-&\hspace*{-10mm}4\eta_1\left[\frac{\eta_1}{1+\eta_1}\{G_C(Q_1^2)
+\frac{\eta_1}{3}G_Q(Q_1^2)-G_M(Q_1^2)\}G_Q(Q_1^2)+\frac{1}{4}G_M^2(Q_1^2)\right],
\end{eqnarray}
and
\begin{equation}
K_{b2}^{ed}(Q_1^2,Q_2^2)=-\int^{x_c(Q_1^2)}_0\frac{dx}{x}b_2^{ed}(x,Q_1^2)
+\int^{x_c(Q_2^2)}_0\frac{dx}{x}b_2^{ed}(x,Q_2^2).
\end{equation}
In Eq.(34), the integral over $x$ should be taken after subtracting the small
$x$ behavior of $b_2^{ed}(x,Q_1^2)$ and $b_2^{ed}(x,Q_2^2)$.
Now, we take $Q_2^2$ large by keeping $Q_1^2$ small or moderate value.
Then,since the Born term at large $Q_2^2$ is negligible, we can neglect it
in Eq.(33). When the integral is convergent, we can take
$x_c(Q_1^2)=x_c(Q_2^2)=0$, hence $K_{b2}^{ed}(Q_1^2,Q_2^2)= 0$.
Then the sum rule (32) relates the tensor polarization 
and the elastic form factors at small or moderate $Q_1^2$ 
to the tensor polarization at large $Q_2^2$. Especially, if
Callan Gross like relation $b_2^{ed}=2xb_1^{ed}\;$ \cite{HJM} with the vanishing tensor
polarization of the sea quark at large $Q_2^2$ holds, the second term on the left-hand
side of Eq.(32) is zero \cite{CK}. In this case the sum rule becomes the
one at small or moderate $Q_1^2$ .
Now the recent experiment at HERMES shows\cite{HERMES,Riedl}
\begin{equation}
\int_{0.002}^{0.85}\frac{dx}{x}b_2^{ed}(x,Q^2=5{\rm GeV}^2) > 0 .
\end{equation}
Though there are unmeasured region, HERMES result possibly suggest the non-zero
tensor polarization at $Q^2=5$GeV$^2$. Since the Born term at
$Q^2=5$GeV$^2$ is negligibly small in Eq.(33), we
can set $B(Q_1^2,Q_2^2)=0$.
Then the sum rule (32) shows that the non-zero polarization persist
in the large $Q^2$ region. 
When the integral over $b_2^{ed}(x,Q^2)/x$ diverges, since the main contribution
comes from the small $x$ region and that ,at large $Q^2$, $b_2^{ed}(x,Q^2)$
behaves similarly as $F_2^{ed}(x,Q^2)$\cite{HJM}, we can expect
$K_{b2}^{ed}(Q_1^2,Q_2^2) > 0$. Thus HERMES result does not
contradict with the zero tensor polarization of the sea quark
at large $Q^2$ in the regularized sense. 
\section{ Summary}
We have derived the sum rules for the structure
functions of the deuteron from the current anti-commutation relation
on the null-plane.
The sum rules correspond to the ones at the wrong signature point.
As explained in the Introduction they give us 
information of the vacuum of the deuteron.\\
From the spin independent part, we obtain the sum rule
for the mean hypercharge of the sea quark of the proton
in the deuteron. Further, in the small $Q^2$ region, 
we obtain the relation among the Born term, the resonances 
and the non-resonant background. From the spin dependent part, 
we obtain the relation between
the tensor polarization at small or moderate $Q^2$ and that
at large $Q^2$. \\
Now, the application of these sum rules in other forms
such as the ones in the photoproduction are possible as in the nucleon
target case. Further, though only the sum rules from the good-good component
are discussed, the same method can be applied to the good-bad component.
In this case, we obtain the sum rules for the spin dependent 
structure functions  $g_1^{ed}$ and $g_2^{ed}$. 
These sum rules take the same form as in the nucleon target case.\cite{K20}\\
\appendix
\section{Current anti-commutation relation on the null-plane through DGS representation}
Let us consider DGS representation of the connected matrix element of 
the current commutation relation on the null-plane between the
stable hadron.\cite{DGS}
\begin{eqnarray}
 W_{ab}(p\cdot q , q^2)& =&\int d^4x \exp (iq\cdot x)<p|[J_a(x) ,J_b(0)]|p>_c\nonumber \\
&=&\int d^4x\exp (iq\cdot x)
\int_0^{\infty}d\lambda^2\int_{-1}^{1}d\beta \exp{(i\beta p\cdot x)}h_{ab}
(\lambda^2,\beta )i\Delta (x,\lambda^2)\nonumber \\
&=&(2\pi )\int_0^{\infty}d\lambda^2\int_{-1}^{1}d\beta \delta ((q+\beta
p)^2-\lambda^2)\epsilon (q^{0}+\beta p^{0})h_{ab}(\lambda^2,\beta ),
\end{eqnarray}
where $W_{ab}(p\cdot q , q^2)$ can be expressed as
\begin{eqnarray}
 W_{ab}(p\cdot q , q^2)& =&\sum_n(2\pi )^4\delta^4(p+q-n)
\langle p|J_a(0)|n\rangle \langle n|J_b(0)|p\rangle \nonumber \\
&-&\sum_n(2\pi )^4\delta^4(p-q-n)
\langle p|J_b(0)|n\rangle \langle n|J_a(0)|p\rangle.
\end{eqnarray}
We denote the lowest mass in the $s$ channel continuum as $M_s$ and
that in the $u$ channel as $M_u$. At the rest frame, $p=(m,\vec{0})$, 
since the first term in Eq.(37) is restricted as $m+q^0=n^0$,
$q^0$ satisfys $ q^0 \geqq M_s -m$. Similarly, since the second term
is restricted as $m-q^0=n^0$, $q^0$ satisfys
$ q^0 \leqq m - M_u$. Hence the first and the second terms
in Eq.(37) are disconnected as far as $m < (M_s + M_u)/2$.\\
\begin{figure}
\centerline{\includegraphics[width=5cm, height=5cm]{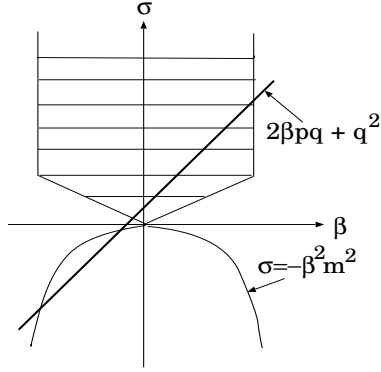}}
\caption{The support property of the spectral
function $h_{ab}(\lambda^2,\beta )$ in the $(\beta ,\sigma )$ plane. It is not zero
only in the shaded region and, below the parabola
$\sigma =-\beta^2m^2$, it is zero by the causality.}
\end{figure}
Now, in the DGS representation, 
$h_{ab}(\lambda^2,\beta )$ is not zero only in the shaded region
in Fig.1.  The integration path is 
$\sigma =2\beta p\cdot q +q^2$, where $\sigma = \lambda^2-\beta^2m^2$.
At the rest frame,
we see that the point in the integration path where the sign changes
through the factor $\epsilon (p\cdot q +\beta m^2)$ lies 
always in the region $\sigma <-\beta^2m^2$.
In the $s$ channel, since $p\cdot q >0$, slope of the integration
path is positive. 
Thus only the region
$\epsilon (p\cdot q +\beta m^2)=1$ contributes,and hence,
in the $s$ channel,we obtain
\begin{eqnarray}
(2\pi )\int_0^{\infty}d\lambda^2\int_{-1}^{1}d\beta \delta ((q+\beta
p)^2-\lambda^2)h_{ab}(\lambda^2,\beta )\theta (q^{0}+\beta p^{0})\nonumber \\
=\sum_n(2\pi )^4\delta^4(p+q-n)
\langle p|J_a(0)|n\rangle \langle n|J_b(0)|p\rangle.
\end{eqnarray}
Similarly, in the $u$ channel, we obtain
\begin{eqnarray}
(2\pi )\int_0^{\infty}d\lambda^2\int_{-1}^{1}d\beta \delta ((q+\beta
p)^2-\lambda^2)h_{ab}(\lambda^2,\beta )\theta (-(q^{0}+\beta p^{0}))\nonumber \\
=\sum_n(2\pi )^4\delta^4(p-q-n)
\langle p|J_b(0)|n\rangle \langle n|J_a(0)|p\rangle.
\end{eqnarray}
By combining these two relations we obtain DGS representation of the
current anti-commutation relation as
\begin{eqnarray}
 \widetilde{W}_{ab}(p\cdot q , q^2)& =& \int d^4x \exp (iq\cdot x)
<p|\{J_a(x) ,J_b(0)\}|p>_c\nonumber \\
&=&\int d^4x\exp (iq\cdot x)\int_0^{\infty}d\lambda^2\int_{-1}^{1}
d\beta\exp{(i\beta p\cdot x)} h_{ab}
(\lambda^2,\beta )\Delta^{(1)} (x,\lambda^2)\nonumber \\
&=&(2\pi )\int_0^{\infty}d\lambda^2\int_{-1}^{1}d\beta \delta ((q+\beta
p)^2-\lambda^2)h_{ab}(\lambda^2,\beta ).
\end{eqnarray}
The null-plane restriction of the current commutation relation or the
anti-commutation relation can be obtained by the integration of $W_{ab}$
and $\widetilde{W}_{ab}$ with respect to $q^-$.

Now we take the scalar current 
$J_a(x)=:\phi ^{\dagger }(x)\tau_a\phi (x):$ where
\begin{equation}
[\phi^{\dagger }(x) , \phi (0)]|_{x^+=0}=i\Delta (x)
\end{equation}
with $\Delta (x)=-\epsilon(x^-)\delta (\vec{x}^{\bot})/4$ at $x^+=0$.
Using this relation, the current commutation relation at $x^{+}=0$ becomes
\begin{equation}
<p|[J_a(x),J_b(0)]|p>_c|_{x^+=0}
= i\Delta (x)<p|:\phi ^{\dagger }(x)\tau_a \tau_b\phi (0): 
+ :\phi ^{\dagger }(0)\tau_b \tau_a\phi (x):|p>_c .
\end{equation}
From Eq.(36) restricted at the null-plane, we have
\begin{equation}
<p|[J_a(x) ,J_b(0)]|p>_c|_{x^+=0}
=\int_0^{\infty}d\lambda^2\int_{-1}^{1}d\beta\exp{(i\beta p\cdot x)} h_{ab}
(\lambda^2,\beta )i\Delta (x,\lambda^2).
\end{equation}
Since  $\Delta (x,\lambda^2)=\Delta (x)=-\epsilon(x^-)\delta (\vec{x}^{\bot})/4$ at $x^{+}=0$, we obtain the
relation
\begin{equation}
<p|:\phi ^{\dagger }(x)\tau_a \tau_b\phi (0): 
+ :\phi ^{\dagger }(0)\tau_b \tau_a\phi (x):|p>_c =
\int_0^{\infty}d\lambda^2\int_{-1}^{1}d\beta\exp{(i\beta p\cdot x)} h_{ab}
(\lambda^2,\beta ).
\end{equation}
By using this relation, the equation restricted 
at the null-plane obtained from Eq.(40) becomes
\begin{eqnarray}
 \lefteqn{<p|\{J_a(x),J_b(0)\}|p>_c|_{x^+=0} }&&\nonumber \\
&=& \int_0^{\infty}d\lambda^2\int_{-1}^{1}d\beta\exp{(i\beta p\cdot x)} h_{ab}
(\lambda^2,\beta )\Delta^{(1)} (x,\lambda^2)\nonumber \\ 
&=&\Delta^{(1)} (x)<p|:\phi ^{\dagger }(x)\tau_a \tau_b\phi (0): 
+ :\phi ^{\dagger }(0)\tau_b \tau_a\phi (x):|p>_c ,
\end{eqnarray}
where we use the fact that
$\Delta^{(1)} (x,\lambda^2)$at $x^{+}=0$ 
is also independent on the mass $\lambda^2$ and 
given as $\Delta^{(1)}(x,\lambda^2)=\Delta^{(1)}(x)=-\ln{|x^-|}\delta(\vec{x}^{\bot})/2\pi$.
A net result is that the current anti-commutation relation on the
null-plane can be obtained from the current commutation relation
on the null-plane simply by changing $i\Delta (x,\lambda^2)$ 
to $\Delta^{(1)}(x,\lambda^2)$ at $x^{+}=0$.
More rigorous reasoning can be done by taking
the Fourier transform and by considering the
conditions necessary to restrict them to the null-plane such as
\begin{equation}
 \lim_{\Lambda\to\infty}\int_{-\infty}^{\infty}dq^-
\exp{(-(q^-)^2/\Lambda^2)}H_{ab}(p\cdot q , q^2),
\end{equation}
where $H_{ab}$ is $W_{ab}$ or $ \widetilde{W}_{ab}$.
This results in the condition to the $h_{ab}$ and 
is known to be equivalent to the superconvergence
relation which is required to get the fixed-mass sum rule,
and in the null-plane formalism is equivalent to
the interchange to set $q^+=0$ and $\nu =p\cdot q$ integration.
It is in this point where the difference between 
the connected matrix element of the stable hadron of the current commutation relation 
and that of the current anti-commutation relation appears.
\section{SU(3) symmetry breaking effect on the symmetry relation}
Here we consider the sum rules for $SU(3)$.
The regularization of the sum rules (19)$\sim$ (22) can be done as
explained in the paragraph before Eq.(19). The detailed method
is given for example in Ref.\cite{K98}. We first summarize the result.
We assume the leading high energy behavior is given by the soft Pomeron as
\begin{equation}
\{F_2^{\bar{\nu}d}+F_2^{\nu d}\} \sim \Big(Q_0^2/Q^2\Big)^{\alpha_P(0)-1}
\beta_{\nu d}(Q^2,1-\alpha_P(0))(2\nu )^{\alpha_P(0)-1},
\end{equation}
where $Q_0^2=1$GeV$^2$ and
\begin{equation}
F_2^{ed} \sim \Big(Q_0^2/Q^2\Big)^{\alpha_P(0)-1}
\beta_{ed}(Q^2,1-\alpha_P(0))(2\nu )^{\alpha_P(0)-1} .
\end{equation}
We expand $\beta_{ld}$ with $l=e$ or $\nu$ as
\begin{equation}
\beta^0_{ld} - (\epsilon -b)\beta_{ld}^1+O((\epsilon -b)^2).
\end{equation}
where the intercept of the Pomeron is set as 
$\alpha_P(0) = 1 + b -\epsilon$ and $\epsilon$ approaches $b$ from the
above. The parameter $\epsilon$ goes to 0 finally after taking out the
pole terms from both-hand side of the sum rule. This change of 
the parameter mimics the $-t$ in the non-forward sum rules. 
Then by assuming that the Pomeron is flavor singlet and that it 
comes from the term $A_0(\alpha ,0)$ being flavor singlet, 
we obtain the relations, $\beta_{\nu d}^0=6\beta_{ed}^0$
and $\beta_{\nu d}^1=6\beta_{ed}^1$ from the sum rules (21) and (22). 
Further from the sum rules (19)
and (21) we obtain the condition 
$\pi\beta_{\nu d}^0=4f_{\pi}^2\beta_{\pi d}$
and from the sum rules (19) and (20) the condition
$f_{\pi}^2\beta_{\pi d}=f_{K}^2\beta_{K d}$.
The regularization dependent terms in the sum rules are related by these
relations and we obtain the relations independent of the regularization.
In this way, we obtain the relation (23) and
\begin{equation}
C_d=\frac{1}{9}(2I_{\pi} + I_{K}),
\end{equation}
where
\begin{equation}
C_d=\int_0^1\frac{dx}{x}\{F_2^{ed}- \beta_{ed}^{0}x^{-b}\} -
 \beta_{ed}^1 .
\end{equation}
Now the condition $f_{\pi}^2\beta_{\pi d}=f_{K}^2\beta_{K d}$
is violated about 20\% phenomenologically. Hence the sum rules
(23) and (27) still diverge if we use the phenomenological value.
To remedy this we consider the mixing of the singlet
and the octet as\cite{K95}
\begin{eqnarray}
\widetilde{A_0}(\alpha ,0)&=& A_0(\alpha ,0)\cos{\theta}
+ A_8(\alpha , 0)\sin{\theta} ,\nonumber \\
\widetilde{A_8}(\alpha ,0)&=& -A_0(\alpha ,0)\sin{\theta}
+ A_8(\alpha , 0)\cos{\theta} ,
\end{eqnarray}
and assume the contribution of the Pomeron is given by the
term $\widetilde{A_0}(\alpha ,0)$. Thus
the residue of the Pomeron has a SU(3) symmetry breaking
piece. Then by rewriting the sum rules (19)$\sim$ (22)
with use of Eq.(52) and by regularizing them, we obtain
the relations 
\begin{equation}
\frac{f_{\pi}^2\beta_{\pi d}}{2(\sqrt{2}\cos{\theta} + \sin{\theta})}
=\frac{f_{K}^2\beta_{Kd}}{2\sqrt{2}\cos{\theta} - \sin{\theta}},
\end{equation}
and
\begin{equation}
\beta_{\nu d}^{i} =
\frac{12(\sqrt{2}\cos{\theta} + \sin{\theta})}{2\sqrt{2}\cos{\theta}
+\sin{\theta}}\cdot \beta_{ed}^{i},
\end{equation}
where $i=0,1$. The relation between $\beta_{\nu d}^0$ and $\beta_{\pi d}$
is the same as the one before the mixing.
In case of the nucleon target, a similar relation as Eq.(53)
was derived, and was found that the relation is satisfied phenomenologically
about at $\theta \sim -13^{\circ}$. In the deuteron case, the relation
(53) seems to be satisfied well phenomenologically about at the same
angle because the large constant term at high energy 
in the cross section formula by the Particle data group\cite{PD} 
satisfys the relation (53) with this angle.  
Now by this mixing, we find that the relation (50) also
holds and the sum rule (23) can be rewritten as
\begin{eqnarray}
\int_{0}^{1}\frac{dx}{x}\left\{\left(\frac{F_2^{\bar{\nu}d}(x,Q^2)+F_2^{\nu d}(x,Q^2)}{2}
\right)-\frac{6(\sqrt{2}\cos{\theta}+\sin{\theta})}{2\sqrt{2}\cos{\theta}+\sin{\theta}}
F_2^{ed}(x,Q^2)\right\}\nonumber \\
= \frac{I_{\pi}^d - I_{K}^d}{3} 
-\frac{3\sin{\theta}}{2\sqrt{2}\cos{\theta}+\sin{\theta}}\cdot C_d  .
\end{eqnarray}
Then the sum rule (27) becomes
\begin{eqnarray}
\lefteqn{(2\sqrt{2}\cos{\theta} - \sin{\theta})I_{\pi}
 -2(\sqrt{2}\cos{\theta} + \sin{\theta})I_K
=6\sqrt{2}(\cos{\theta} - \sqrt{2}\sin{\theta})}&&\nonumber \\
&+& \int_0^1dx\Big\{ 4\sqrt{2}(\cos{\theta} -
\sqrt{2}\sin{\theta})(\lambda_u + \lambda_d)
-8(\sqrt{2}\cos{\theta} + \sin{\theta})\lambda_s\Big\} . 
\end{eqnarray}
Since the strange sea quark is suppressed above $x=0.01$ greatly,
the large symmetry restoration of the sea quark
exists in the small $x$ region\cite{K03}, 
and that to satisfy Eq.(56)
the small $x$ limit of the strange sea quark distribution must 
be larger than that of the
$u$ or $d$ type sea quark.
\section{The Born term contributions}
The Born term contribution to Eq.(1) is given as
\begin{equation}
\frac{1}{2}\delta (2\nu - Q^2)B^{\mu\nu},
\end{equation}
with 
\begin{eqnarray}
B^{\mu\nu}=
 <p,E|J^{\mu}_{em}(0)|n,E^{\prime}><n,E^{\prime}|J^{\nu}_{em}(0)|p,E>\nonumber\\
=\frac{1}{m_d^4}\sum_{\lambda}\Big(\big\{G_1(Q^2)(E^{*}\cdot E^{\prime})
 -G_3(Q^2)\frac{(E^{*}\cdot q)
(E^{\prime}\cdot q)}{2m_d^2}\big\}(p+n)^{\mu} \nonumber \\
+ G_M(Q^2)\big\{-E^{\prime \mu}(E^{*}\cdot q)+E^{*
 \mu}(E^{\prime}\cdot q)\big\}\Big)\nonumber\\
\times \Big(\big\{G_1(Q^2)(E^{\prime *}\cdot E)
 -G_3(Q^2)\frac{(E^{\prime *}\cdot q)
(E\cdot q)}{2m_d^2}\big\}(p+n)^{\nu} \nonumber\\
+ G_M(Q^2)\big\{E^{\nu}(E^{\prime *}\cdot q)-E^{\prime * \nu}
(E\cdot q)\big\}\Big),
\end{eqnarray}
where  $n=p+q$ and $\lambda$ is the polarization of $E^{\prime}$.
Here we denote it as $E^{\prime}(n,\lambda )$.
Then using the relation
\begin{equation}
\sum_{\lambda}E^{\prime * \mu}(n,\lambda )E^{\prime \nu}(n,\lambda )
= n^{\mu}n^{\nu} - m_d^2g^{\mu\nu}
\end{equation}
with $E^{\prime}\cdot n=E^{* \prime}\cdot n=0$ and
$E^{\prime}\cdot E^{*\prime}=-m_d^2$,
we take the product on the right-hand side of Eq.(58).
Then we classify the product into the symmetric terms and the
antisymmetric ones under
the interchange of the $E$ and $E^{*}$. We first calculate
the symmetric ones and take the polarization average of the
initial deuteron and obtain the Born term contribution
to $F_1$ and $F_2$ as
\begin{equation}
F_1 = \delta (2\nu - Q^2)\frac{Q^2}{3}(\eta + 1)G_M^2,
\end{equation}
and
\begin{equation}
F_2 = \delta (2\nu - Q^2)Q^2\Big( G_C^2
+\frac{8}{9}\eta^2G_Q^2 + \frac{2}{3}\eta G_M^2\Big).
\end{equation}
The rest of the symmetric terms contribute to $b_1 \sim b_4$.
By noting that the polarization averaged parts are subtracted,
and that  $g^{\mu\nu}$ is only in the tensor $G^{\mu\nu}$
we find the contribution to the sum of the $b_1$ and the
$b_2 - 3 b_3$ with an appropriate coefficient. 
Further since $E^{\mu}E^{*\nu}+E^{*\mu}E^{\nu}$
is only in the tensor $u^{\mu\nu}$, we obtain the
contribution to the $b_2 - 3 b_3$. Hence we can
separate the contribution to the $b_1$ and the $b_2 - 3 b_3$. 
Similar consideration can be done to the coefficient of
$t^{\mu\nu}$ which gives the contribution to the $b_2-3b_4$ and
$p^{\mu}p^{\nu}$ which gives the contribution
to the $b_2 + 3b_3 + 3b_4$. Thus we obtain 
\begin{equation}
b_1=\delta (2\nu - Q^2)\frac{Q^2}{2}\eta G_M^2,
\end{equation}
\begin{equation}
b_2= \delta (2\nu - Q^2)4Q^2\eta^2\Big(\frac{1}{1+\eta}\big(G_C
+\frac{\eta}{3}G_Q-G_M\big)G_Q+\frac{1}{4\eta}G_M^2\Big),
\end{equation}
\begin{equation}
b_3=\delta (2\nu - Q^2)4Q^2\eta^2\Big(\frac{1}{3(1+\eta)}\big(G_C
+\frac{\eta}{3}G_Q-G_M\big)G_Q-\frac{3\eta+2}{12\eta}G_M^2\Big),
\end{equation}
\begin{equation}
b_4=\delta (2\nu - Q^2)4Q^2\eta^2\Big(\frac{1}{3(1+\eta)}\big(G_C
+\frac{\eta}{3}G_Q-G_M\big)G_Q\nonumber\\
+\frac{1+6\eta}{12\eta}G_M^2+G_QG_M\Big).
\end{equation}
Now the antisymmetric parts under the interchange of the $E$ and $E^{*}$
give the contribution to the $g_1$ and $g_2$.
In this case, we first note the identity
\begin{eqnarray}
\lefteqn{a^{\mu}\epsilon^{\nu\rho\alpha\beta}q_{\rho}s_{\alpha}p_{\beta}
-a^{\nu}\epsilon^{\mu\rho\alpha\beta}q_{\rho}s_{\alpha}p_{\beta}}&&\nonumber\\
&=&-(a\cdot q)\epsilon^{\mu\nu\alpha\beta}s_{\alpha}p_{\beta}
-(a\cdot s)\epsilon^{\mu\nu\alpha\beta}p_{\alpha}q_{\beta}
+(a\cdot p)\epsilon^{\mu\nu\alpha\beta}s_{\alpha}q_{\beta}.
\end{eqnarray}
Since 
$s^{\mu}=-(i/m_d^2)\epsilon^{\mu\alpha\beta\gamma}E^{*}_{\alpha}E_{\beta}p_{\gamma}$,
we have
\begin{equation}
\epsilon^{\nu\rho\alpha\beta}q_{\rho}s_{\alpha}p_{\beta}
=i(E^{*\nu}(q\cdot E) - E^{\nu}(q\cdot E^{*})) .
\end{equation}
Thus we obtain the relation
\begin{eqnarray}
\lefteqn{i\Big(a^{\mu}\big(E^{*\nu}(q\cdot E) - E^{\nu}(q\cdot E^{*})\big)
-a^{\nu}\big(E^{*\mu}(q\cdot E) - E^{\mu}(q\cdot E^{*})\big)\Big)}&&\nonumber\\
&=&-(a\cdot q)\epsilon^{\mu\nu\alpha\beta}s_{\alpha}p_{\beta}
-(a\cdot s)\epsilon^{\mu\nu\alpha\beta}p_{\alpha}q_{\beta}
+(a\cdot p)\epsilon^{\mu\nu\alpha\beta}s_{\alpha}q_{\beta}.
\end{eqnarray}
In case of $a=2p+q$, $a\cdot q=2p\cdot q + q^2 =0$ for the Born term.
Another useful relation is
\begin{equation}
\epsilon^{\mu\nu\alpha\beta}s_{\alpha}p_{\beta}
=i(E^{*\mu}E^{\nu}-E^{*\nu}E^{\mu}).
\end{equation}
In this way we obtain
\begin{equation}
g_1=\delta (2\nu -Q^2)\frac{Q^2}{2} G_M(G_C + \frac{\eta}{3}G_Q+\frac{\eta}{2}G_M),
\end{equation}
\begin{equation}
g_2=\delta (2\nu -Q^2)\frac{Q^2\eta}{2}G_M(G_C + \frac{\eta}{3}G_Q - \frac{1}{2}G_M).
\end{equation}

\end{document}